%% file: main.tex
\documentclass[conference]{IEEEtran}
\IEEEoverridecommandlockouts
\usepackage{cite}
\usepackage{amsmath,amssymb,amsfonts}
\usepackage{algorithmic}
\usepackage{graphicx}
\usepackage{textcomp}
\usepackage[table]{xcolor}
\usepackage{comment}
\usepackage{multirow}
\usepackage{booktabs}
\usepackage{siunitx}
\def\BibTeX{{\rm B\kern-.05em{\sc i\kern-.025em b}\kern-.08em
    T\kern-.1667em\lower.7ex\hbox{E}\kern-.125emX}}
\begin{document}

\title{Preserving Speech-to-Text LLM Capabilities in Speech-to-Speech Generation\\
}
\author{\IEEEauthorblockN{Yuxuan Hu, Heng Lu, Ruchao Fan, Yao Qian, Xiaofei Wang, Jian Xue, Heming Wang, \\
Shuohang Wang, Young Jin Kim, Yelong Shen, Jinyu Li}
\IEEEauthorblockA{Microsoft, USA}}

\maketitle

\begin{abstract}
Strong speech-to-text (S2T) LLMs already provide robust speech perception and text reasoning, but adding speech-to-speech (S2S) output is challenging: fine-tuning the backbone can degrade the original S2T performance, while attaching a downstream talker reintroduces a text-to-speech bottleneck. We present PRIME-Speech, a frozen-backbone S2S conversion framework that trains only speech-generation modules. PRIME-Speech synchronizes a causal audio post-decoder with intermediate hidden states of the frozen backbone, so codec tokens are generated from the model's evolving reasoning trajectory rather than from completed text chunks. The post-decoder uses mixed hidden-state, text, and audio-history conditioning, and a training-time packing strategy with turn-level audio KV-cache and position reset stabilizes multi-turn spoken interaction without additional multi-turn S2S training data. Multi-token prediction further reduces the effective codec prediction rate and improves first-audio latency without modifying the reasoning path. Across speech translation, spoken QA, speech understanding, and multi-turn dialogue, PRIME-Speech preserves the S2T behavior of the frozen backbone while producing accurate, low-WER spoken responses.

\end{abstract}


\input{sections/introduction}

\input{sections/method}
\input{sections/training}
\input{sections/experiments}

\input{sections/conclusion}

\bibliographystyle{IEEEtran}
\bibliography{mybib}
\section{Generative AI Use Disclosure}
We used generative AI tools only for language polishing, including grammar correction and spelling checks, and for assisting in formatting and drawing LaTeX tables. All technical content, experiments, and conclusions were produced and verified by the authors.
\end{document}

%% file: sections/introduction.tex
\section{Introduction}
\label{sec:introduction}

Speech-to-speech (S2S) interaction allows language assistants to communicate with users naturally: a system listens to speech, reasons about the user's intent, and responds directly in speech. The dominant approach remains a cascade, where automatic speech recognition (ASR) converts speech into text, a text LLM generates a response, and a text-to-speech (TTS) system converts it back into speech~\cite{cui2025recent, arora2025landscape, wang2023whislu, ling2024adapting}. Although modular and convenient, this pipeline passes recognition errors to the LLM and cannot generate speech until enough text is available. These limitations motivate S2S models that integrate speech generation with LLM decoding rather than treating it as a separate post-processing step.

This paper studies how to convert a strong speech-to-text (S2T) LLM into an S2S model without losing the capabilities of the original backbone. Modern S2T LLMs already support speech perception, text reasoning, and instruction following~\cite{zhang2023speechgpt, chu2024qwen2, phi4-mini}. While prior work has studied the gap between S2T and text-based language modeling~\cite{xiang2025understanding, fan2025alignformer, cuervo2025closing, wang2025speech, wang2026closing}, the challenges of extending S2T LLMs to generate speech remain less understood. Adding text-to-speech synthesis may seem sufficient~\cite{eskimez2024e2, chen2025neural, du2024cosyvoice, du2025cosyvoice, yang2025pseudo, zhou2026indextts2, liao2026fish, zhu2026zipvoice}, but end-to-end S2S requires more than synthesizing a completed response. Generating speech tokens within the same autoregressive process as text generation can disrupt the backbone's learned behavior and weaken its speech understanding, reasoning, and instruction following. If speech output begins only after the text response is complete, the system loses streaming and brings back cascade latency. S2S adaptation must therefore preserve the backbone's capabilities and keep speech generation synchronized.

Existing S2S architectures reflect this trade-off~\cite{defossez2024moshi, wang2024freeze, zhang2025mimo, ding2025kimi, chen2025slam, chen2025fun, yang2026moss}. Unified token-interleaving models generate text and audio in a single autoregressive stream. This keeps the two outputs synchronized but requires the main decoder to learn different text and audio objectives~\cite{zeng2024glm,fang2025llama, wu2025step}. Thinker--Talker systems separate reasoning from speech generation~\cite{xu2025qwen2,xu2025qwen3,chen2025minmo,team2026qwen3}. However, the talker usually receives completed text, text chunks, or hand-off states from the text branch, not the backbone's evolving hidden states. Multi-token prediction (MTP) has become an effective way to accelerate text generation~\cite{liu2024deepseek,gloeckle2024better}. Applying it to speech can also reduce autoregressive codec decoding steps~\cite{long2025vita,wang2025vocalnet}, but predicting multiple speech tokens at once often degrades generation quality.

\begin{figure*}[t]
    \centering
    \includegraphics[width=0.85\textwidth]{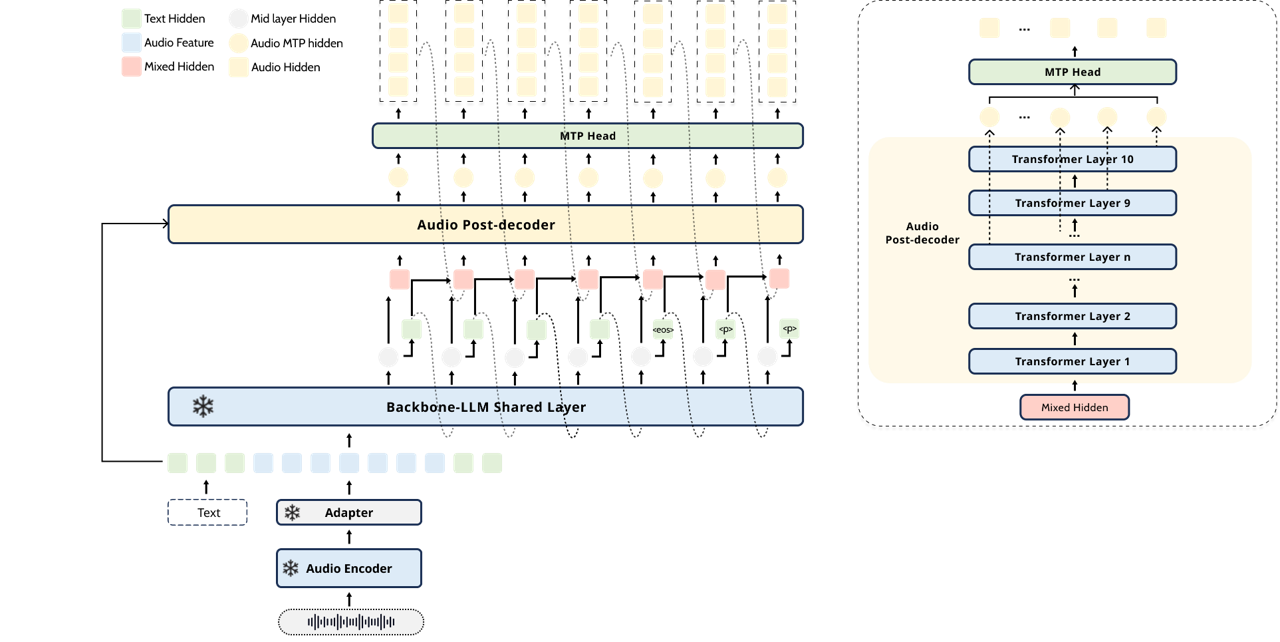}
    \caption{Overview model architecture of PRIME-Speech. The frozen S2T backbone retains its original text reasoning path, while a parallel audio post-decoder generates speech from intermediate hidden states. Layer-wise MTP uses successive post-decoder states to predict multiple codec tokens per update.}
    \label{fig:arch}
\end{figure*}

We propose PRIME-Speech, a framework for \textbf{P}reserving \textbf{R}easoning and \textbf{I}ntelligence while enabling \textbf{M}ore \textbf{E}fficient Speech-to-Speech generation. Its key idea is \emph{hidden-state synchronization}, as shown in Fig.~\ref{fig:arch}. At each decoding step, PRIME-Speech takes the hidden state from an intermediate layer of the frozen backbone. From this point, text and audio follow separate paths in parallel. The text path continues through the remaining backbone layers and the original text head, leaving the S2T computation unchanged. The audio path feeds the same intermediate state into a causal audio post-decoder whose depth matches the remaining backbone layers. This gives the audio branch access to the backbone's current reasoning state before the text response is complete. The intermediate state also provides a richer conditioning signal for speech generation than a discrete text token.

The audio post-decoder further supports multi-token prediction through layer-wise audio heads. For the $k$-th prediction, the corresponding audio head takes the hidden state from the $k$-th layer counted backward from the end of the post-decoder. The audio heads therefore draw on hidden states from different post-decoder layers. This layer-wise design propagates contextual information across the codec tokens predicted within the same update. As a result, each synchronized update produces multiple codec tokens while evaluating the shared backbone only once. One backbone computation therefore covers more speech output, reducing the number of decoding steps and the end-to-end generation cost. All MTP operations remain on the audio path, so the original text reasoning path is unchanged.

We evaluate PRIME-Speech from four aspects: (1)~preservation of the backbone's S2T capabilities; (2)~the accuracy of spoken responses and their consistency with the corresponding text responses; (3)~stability in multi-turn dialogue; and (4)~the effect of audio-side MTP on latency and real-time factor. Experiments cover multilingual S2S translation, spoken question answering, speech understanding, and multi-turn dialogue. The results show that PRIME-Speech largely preserves the original S2T performance while generating accurate spoken responses with low rendering WER. Its hybrid cache policy supports stable multi-turn interaction, and audio-side MTP substantially improves decoding efficiency. Our main contributions are:

\begin{itemize}
    \item We introduce \emph{hidden-state synchronization} for converting a pretrained S2T LLM into an S2S model. A frozen text pathway preserves the original S2T computation, while a parallel speech branch follows the backbone's intermediate reasoning states.
    \item We develop a causal audio post-decoder that runs concurrently with the remaining text layers. It conditions on the backbone's intermediate state, text embeddings, and recent audio history during speech generation.
    \item We propose a hybrid cache policy for multi-turn dialogue. The text cache is retained across turns to preserve dialogue context, while the audio cache is reset at the start of each turn to limit acoustic drift.
    \item We introduce a layer-wise audio MTP design that uses hidden states from successive post-decoder layers to maintain context across codec tokens predicted in the same update. It reduces the real-time factor by more than $3\times$ with stable task performance and no change to the text reasoning path.

\end{itemize}

%% file: sections/method.tex
\section{Method}
\label{sec:method}

\subsection{Overview}
\label{sec:method_overview}
PRIME-Speech converts a frozen S2T LLM into an S2S model by adding a trainable speech-generation branch around the original text pathway. Let \(x=(x^\tau,x^a)\) denote the input, where \(x^\tau\) is an optional text prompt and \(x^a\) is the input speech waveform. The frozen backbone produces a text response \(y^\tau\), while the added audio branch produces a codec-token response \(y^a\). The conversion is written as
\begin{equation}
P(y^\tau, y^a \mid x)
= P_{\mathrm{bb}}(y^\tau \mid x)\,
  P_{\mathrm{aud}}(y^a \mid y^\tau, H^{\mathrm{mid}}; \theta_a),
\label{eq:factorization}
\end{equation}
where \(P_{\mathrm{bb}}\) is the frozen backbone distribution, \(P_{\mathrm{aud}}\) is the trainable audio-branch distribution, \(H^{\mathrm{mid}}\) is a sequence of intermediate backbone states, and \(\theta_a\) denotes the audio-branch parameters. During streaming, the dependence on \(y^\tau\) is prefix-restricted. At decoding update \(s\), the frozen backbone exposes a hidden state \(h^{\mathrm{mid}}_s\); the text head and audio branch consume this state in parallel. The audio branch uses only the text and audio history committed before \(s\), while the text token emitted at \(s\) becomes available to the audio branch at the next update. Thus the backbone remains responsible for speech perception and reasoning, and the audio branch acts as a synchronized speech readout of the preserved backbone trajectory.

\subsection{Frozen Backbone and Codec Targets}
\label{sec:backbone_codec}
The backbone maps speech and text inputs into a shared autoregressive context. For an input waveform \(x^a\), the frozen speech encoder and projection module produce acoustic embeddings \(e^{\mathrm{sp}}\); text tokens \(x^\tau\) are mapped to embeddings \(e^\tau\). The frozen transformer stack processes the concatenated sequence,
\begin{equation}
h^{(\ell)}
=\mathrm{Backbone}^{(\ell)}([e^\tau,e^{\mathrm{sp}}]),
\quad \ell=1,\ldots,L .
\label{eq:layerwise_hidden}
\end{equation}

We use a fixed middle-layer stream \(H^{\mathrm{mid}}=h^{(\ell_{\mathrm{mid}})}\) as the speech-conditioning interface, with \(\ell_{\mathrm{mid}}\) set to the layer at approximately two-thirds depth of the backbone. This choice is guided by a layer-wise centered kernel alignment (CKA) analysis~\cite{kornblith2019similarity}, which identifies this depth as carrying the richest paralinguistic information while retaining semantic grounding. The final text logits, text-token embeddings, and text KV cache remain those of the frozen backbone, so the original S2T pathway is not updated by S2S training.

Target speech is represented by semantic codec tokens from the CosyVoice2~\cite{du2024cosyvoice} tokenizer at 25\,Hz:
\begin{equation}
y^a=\{y^a_t\}_{t=1}^{T_a}, \quad
y^a_t\in\{1,\ldots,V_a\},
\label{eq:codec_tokens}
\end{equation}
where \(T_a\) is the number of codec frames and \(V_a\) is the codec vocabulary size. The audio post-decoder predicts these tokens, and the paired codec decoder converts them back to waveform.

\subsection{Hidden-State-Synchronized Audio Post-Decoder}
\label{sec:post_decoder}
The audio post-decoder is a causal transformer executed in the same streaming update loop as the frozen text path. We index this loop by \(s\). At update \(s\), the frozen backbone state \(h^{\mathrm{mid}}_s\) fans out to two branches: the text head predicts one text token \(y^\tau_s\), and the audio post-decoder predicts an audio block \(\mathbf{y}^a_s=(y^a_{s,1},\ldots,y^a_{s,B_s})\). The flat codec sequence in Eq.~\eqref{eq:codec_tokens} is viewed as consecutive blocks in this loop; \(B_s=1\) before MTP and \(B_s\le k\) after MTP is enabled. The two branches are concurrent, but the audio branch is causal: it conditions on the current hidden state and the history committed before \(s\), not on the text token being predicted in the same update. The post-decoder therefore models
\begin{equation}
P(\mathbf{y}^a_s \mid \mathbf{y}^a_{<s}, y^\tau_{<s},
H^{\mathrm{mid}}_{\le s}; \theta_a).
\label{eq:audio_ar}
\end{equation}
After both branches emit their outputs, \(y^\tau_s\) and \(\mathbf{y}^a_s\) are committed and become the text and audio history for update \(s{+}1\). Thus synchronization is timestamp-level rather than chunk-level: PRIME-Speech does not wait for completed text, fixed text chunks, or word-to-frame force alignment, and the audio branch does not run on a free-standing clock detached from the backbone trajectory.

The conditioning state combines three signals: the synchronized backbone state, the previous text embedding, and recent audio history. Let \(e^\tau_{s-1}\) be the embedding of the previously committed text token and
\begin{equation}
r^a_{s-1}=\frac{1}{B_{s-1}}\sum_{j=1}^{B_{s-1}} e^a_{s-1,j}
\label{eq:audio_summary}
\end{equation}
be the mean embedding of the codec tokens committed by the previous audio update; beginning-of-sequence embeddings are used for \(s=1\). The mixed conditioning vector is
\begin{equation}
h^{\mathrm{mix}}_s
= w_h h^{\mathrm{mid}}_s
 + w_\tau e^\tau_{s-1}
 + w_a r^a_{s-1},
\label{eq:mixed_state}
\end{equation}
with \(w_h=w_\tau=w_a=1.0\) in our experiments, selected as the best simple fixed-weight setting in held-out ablations. The three terms provide semantic state, lexical anchoring, and local acoustic continuity, respectively. Text and audio are therefore generated by parallel branches rather than by a single interleaved token stream: they maintain separate caches, and hidden states serve as the synchronization interface.

Training follows the same causal graph. With teacher forcing, the frozen backbone is evaluated on the reference text prefix and the audio post-decoder receives reference histories up to \(s{-}1\); only the current audio block contributes to the speech-generation loss. At inference, generated text tokens and audio blocks replace the reference histories. In both cases, the audio branch never conditions on future text or on \(y^\tau_s\), the text token generated concurrently at update \(s\).


\subsection{Multi-Token Prediction for Codec Efficiency}
\label{sec:mtp}
Codec tokens are generated at 25\,Hz, so autoregressive audio decoding can dominate response latency. After the single-token audio branch has learned stable hidden-state alignment, PRIME-Speech attaches MTP heads to the audio post-decoder. In the timestamp formulation, MTP sets the audio block size. At update \(s\), one post-decoder state predicts \(k\) future single-codebook codec-token distributions:
\begin{equation}
p_{s,i}
=P(y^a_{s,i}\mid \mathbf{y}^a_{<s}, y^\tau_{<s},
H^{\mathrm{mid}}_{\le s};\theta_a),
\quad i=1,\ldots,k .
\label{eq:mtp_pred}
\end{equation}
The objective is a weighted sum of valid future-token losses,
\begin{equation}
\mathcal{L}_{\mathrm{mtp}}
=-\sum_s\sum_{i=1}^{k}
\lambda_i\log p_{s,i}(y^a_{s,i}),
\label{eq:mtp_loss}
\end{equation}
where positions beyond the utterance boundary are masked. At inference, each synchronized audio update commits up to \(k\) codec tokens, reducing the effective codec prediction rate from 25\,Hz to \(25/k\)\,Hz. MTP therefore accelerates the PRIME audio branch without modifying the frozen reasoning pathway.

\subsection{Training-Time Multi-Turn Packing and Cache Reset}
\label{sec:multiturn}
Realistic multi-turn S2S data is costly to collect, and training with long cross-turn audio histories is unstable. PRIME-Speech therefore does not require additional multi-turn S2S supervision for this component. Instead, it reuses the frozen S2T backbone's existing ability to maintain dialogue-level text context and trains the audio branch with a turn-level cache policy. During training, we concatenate unrelated single-turn examples into packed pseudo-dialogues. This exposes the audio branch to long text-side context and explicit turn boundaries, but it prevents the model from treating the previous turn's acoustic realization as useful context for the next turn.

The policy is applied consistently in training and inference. Text KV states are accumulated across turns, while audio KV states are reset for each assistant turn. Let \(n\) index either a real dialogue turn at inference time or a packed single-turn segment during training, and let \(m\in\{\tau,a\}\) denote text or audio modality. The cache update is
\begin{equation}
\mathbf{C}^{(n)}_m=
\begin{cases}
\mathbf{C}^{(<n)}_\tau
\oplus\{\mathbf{K}^{(n)}_\tau,\mathbf{V}^{(n)}_\tau\},
& m=\tau,\\
\{\mathbf{K}^{(n)}_a,\mathbf{V}^{(n)}_a\},
& m=a,
\end{cases}
\label{eq:cache_update}
\end{equation}
where \(\oplus\) denotes concatenation and \(\mathbf{C}^{(<n)}_\tau\) stores text states from preceding turns. Codec prediction in turn \(n\) is therefore conditioned on accumulated text semantics and turn-local audio history:
\begin{equation}
P(y^a_t\mid \mathbf{C}^{(<n)}_\tau,\mathbf{C}^{(n)}_a,
H^{\mathrm{mid}};\theta_a).
\label{eq:hybrid_cond}
\end{equation}

The same boundary is also applied to audio positions. When packed single-turn examples are used for training, the text positions remain global in the packed sequence, but the audio positional index is reset at the beginning of every packed segment. At inference, the same reset is applied at each new assistant response. If \(i\) is a token index, \(m_i\) its modality tag, and \(s_n\) the starting index of the current audio segment, the position used by the audio branch is
\begin{equation}
\mathcal{P}^{(n)}(i)=
\begin{cases}
i, & m_i \in \mathrm{Text},\\
i-s_n, & m_i \in \mathrm{Audio}_n .
\end{cases}
\label{eq:pos_reset}
\end{equation}
This training-time packing plus turn-local audio reset lets PRIME-Speech exploit the backbone's text-side multi-turn capability without collecting new multi-turn S2S data, while preventing stale audio states from causing repetition or drift.

%% file: sections/training.tex
\section{Experimental Setup}
\label{sec:setup}

\subsection{Training Data}
\label{sec:data}
Our training mixture comprises speech transcription, multilingual speech translation, spoken question answering, and assistant-style response data. Each example contains paired text and speech targets without token-level forced alignment. The two streams advance concurrently at a \(1:k\) ratio, where \(k\) denotes the MTP rate: each step generates one text token and \(k\) audio codec tokens. Given the higher information density and faster decoding rate of text, the text stream reaches its end before the audio stream. Text-side padding then maintains synchronized decoding until speech generation is complete, ensuring that audio does not advance beyond the available text context. After task balancing and resampling, the mixture contains about 100k weighted hours, as summarized in Table~\ref{tab:dataset_specs}.

\input{tables/datasets}

We use transcribed English speech from LibriHeavy~\cite{kang2023libriheavy} as the primary reconstruction data. These examples provide dense supervision for codec prediction and help stabilize the randomly initialized audio post-decoder during early training.

To train the audio branch beyond transcript reconstruction, we include multilingual speech-to-English translation data. We use CVSS~\cite{jia2022cvss} with a subset of CoVoST-2 X2EN~\cite{wang2021covost2} covering the seven source languages used in~\cite{hu2025slms2st}. Because public S2ST data are limited, we also synthesize about 10k hours of X-to-English speech targets using the data generation and filtering procedure described in~\cite{hu2025slms2st}.

We include VoiceAssistant-400K~\cite{xie2024mini}, TriviaQA~\cite{2017arXivtriviaqa}, and Natural Questions\footnote{https://huggingface.co/datasets/sentence-transformers/natural-questions}. For TriviaQA and Natural Questions, we follow the synthesis and processing procedure of~\cite{wu2025towards}. This subset exposes the post-decoder to answer-like outputs, including short factual answers and longer explanatory responses.

For all components, target-side speech is synthesized from text using Microsoft Azure TTS with speakers sampled from several hundred identities, and the waveforms are encoded by the CosyVoice2~\cite{du2024cosyvoice} tokenizer at 25\,Hz. This controlled target-speech construction gives clean supervision for semantic alignment and intelligibility. Accordingly, our evaluation focuses on transcript-level task correctness, S2T--S2S consistency, WER, and decoding efficiency, and reports the UTMOS-based fluency score from VocalBench as a speech-naturalness indicator, rather than claiming improvements in expressive prosody. 
\input{tables/overview_results}
\subsection{Model Configuration}
\label{sec:model_training}
PRIME-Speech uses Phi-4-MM-7B as its backbone, a 7B-parameter S2T LLM pre-trained on 2M hours of speech and 5T text tokens. Starting from this backbone allows us to study whether speech generation can be added without degrading its existing speech understanding and text reasoning capabilities. This effect is difficult to isolate in released S2S models~\cite{xu2025qwen2, ding2025kimi, wu2025step}, where the model architecture, training data, and speech-generation fine-tuning are already coupled.

\subsubsection{Trainable modules}
The audio post-decoder consists of 10 causal Transformer layers with approximately 2B parameters. Its hidden size and attention configuration match those of the backbone, making the intermediate representation compatible with the speech branch. The layer-wise MTP component contains multiple MLP prediction heads with a hidden dimension of 2048 and approximately 100M parameters.

\subsubsection{Frozen modules}
The speech encoder, projection module, and text LLM backbone retain their pre-trained parameters. This confines adaptation to the speech-generation components and leaves the original S2T computation unchanged.

\subsection{Training Curriculum}
\label{sec:curriculum}
Training proceeds in two stages. The first stage trains the audio post-decoder with standard next-token codec prediction, jointly covering hidden-state-to-codec alignment and semantic rendering. The second stage enables multi-token codec prediction. This ordering avoids asking a randomly initialized speech branch to solve long-horizon codec prediction before it can reliably follow the frozen backbone states.

\subsubsection{Audio-branch training}
Stage 1 trains the audio post-decoder at the native 25\,Hz codec rate with standard next-token prediction. We train on the full task-balanced mixture for one epoch using AdamW, a learning rate of \(1\times10^{-4}\), and linear decay. LibriHeavy provides dense audio-text correspondence for stable acoustic-token modeling, while translation, spoken QA, and assistant-style examples teach the post-decoder to render content inferred or generated by the frozen backbone rather than only verbatim transcripts. Because the backbone is frozen, this stage improves speech realization rather than changing the model's reasoning ability.

\subsubsection{MTP training}
Stage 2 enables MTP and continues training for 20k steps with the same learning rate schedule. ASR data are down-weighted, while translation and spoken QA data remain active. At this point the post-decoder has already learned stable hidden-state conditioning, so MTP is trained as an efficiency component that compresses codec-token prediction without modifying the frozen reasoning path.

%% file: tables/datasets.tex
\begin{table}[t]
\centering
\footnotesize
\setlength{\tabcolsep}{4pt}
\renewcommand{\arraystretch}{0.85}
\caption{Statistics of datasets used for curriculum training.}
\label{tab:dataset_specs}
\resizebox{\columnwidth}{!}{%
\begin{tabular}{l l c r c}
\toprule
Dataset & Type & Lang. & Hrs & Stage \\
\midrule
LibriHeavy \cite{kang2023libriheavy} & TTS & EN & 46k & S1 \\
In-house X2EN & S2ST & EN & 10k & S1, S2 \\
CoVoST-2 X2EN \cite{wang2021covost2} & S2ST & EN & 1k & S1, S2 \\
VoiceAssistant~\cite{xie2024mini} & SQA & EN & 4k & S1, S2 \\
TriviaQA \cite{wu2025towards} & SQA & EN & 2k & S1, S2 \\
\bottomrule
\end{tabular}
}
\end{table}

%% file: tables/overview_results.tex
\begin{table*}[t]
\renewcommand{\arraystretch}{1.03}
\setlength{\tabcolsep}{3.2pt}
\centering
\caption{Performance comparison across speech tasks. For entries reported as ``S2T / S2S'', the left number is the S2T score and the right number is the S2S score. ``Flu.'' denotes the UTMOS score in VocalBench. ``--'' indicates unavailable results.}
\label{tab:speech_results_transposed}
\resizebox{\textwidth}{!}{
\begin{tabular}{l cc cccc ccc cccccc cc}
\toprule
& \multicolumn{2}{c}{S2ST}
& \multicolumn{7}{c}{Speech Conversation}
& \multicolumn{8}{c}{Speech Understanding} \\
\cmidrule(lr){2-3}\cmidrule(lr){4-10}\cmidrule(lr){11-18}
Model
& \multicolumn{2}{c}{X2EN}
& \multicolumn{4}{c}{UltraEval-Audio}
& \multicolumn{3}{c}{Multi-turn}
& \multicolumn{6}{c}{VocalBench}
& \multicolumn{2}{c}{BigBench-Audio} \\
\cmidrule(lr){2-3}\cmidrule(lr){4-7}\cmidrule(lr){8-10}\cmidrule(lr){11-16}\cmidrule(lr){17-18}
& FLEURS & CoVoST
& LLaMA-QA & TriviaQA & WebQ & WER$\downarrow$
& S2T & S2S & WER$\downarrow$
& Know. & Reas. & Creat. & Flu. & Single & Overall
& S2T & S2S \\
\midrule

Qwen3-Omni-30B
& 33.25 / 32.72 & 41.25 / 37.62
& 83.00 / 71.33 & 61.43 / 57.52 & 55.95 / 52.51 & 14.92
& 79.89 & 70.39 & 11.28
& 89.4 & 4.43 & 4.77 & 4.38 & 4.96 & 91.99
& 83.7 & 72.0 \\

GPT-4o
& 33.86 / - & 37.09 / -
& 83.00 / - & 76.07 / - & 50.98 / - & --
& -- & -- & --
& 91.3 & 4.69 & 3.93 & 4.16 & 4.67 & 88.06
& 70.2 & 67.2 \\
\midrule
GLM-4-Voice-9B
& -- & --
& 64.70 / 50.70 & 39.10 / 26.50 & 32.20 / 15.90 & --
& 74.86 & 70.95 & 7.83
& 56.4 & 3.64 & 3.29 & 3.87 & 3.62 & 68.94
& 44.8 & 42.7 \\

Kimi-Audio-7B
& 7.68 / - & 7.40 / -
& 76.67 / 62.33 & 46.78 / 37.99 & 41.98 / 35.37 & 14.85
& 73.18 & 65.36 & 10.9
& 62.2 & 3.13 & 3.10 & 2.36 & 3.15 & 59.38
& 59.4 & 51.0 \\

Step-Audio-2-Mini-7B
& 29.03 / 24.85 & 33.25 / 27.21
& 61.00 / 60.33 & 33.40 / 32.23 & 33.02 / 31.69 & 8.56
& 70.39 & 69.27 & 6.15
& 58.5 & 3.67 & 3.12 & \textbf{4.52} & 3.44 & 70.71
& 50.9 & 47.5 \\

VocalNet-8B$^\dagger$
& -- & --
& 76.33 / 69.00 & 44.63 / 38.38 & \textbf{44.05} / 39.27 & 7.68
& 74.86 & 68.72 & 8.52
& 68.0 & 3.75 & 3.51 & 4.45 & 3.53 & 74.54
& 45.9 & 44.9 \\

Qwen2.5-Omni-7B
& \textbf{34.59} / 5.94 & 39.72 / 10.52
& 76.33 / 71.00 & \textbf{47.66} / \textbf{45.60} & 42.18 / 39.42 & 21.5
& 69.83 & 67.04 & 4.23
& \textbf{69.5} & \textbf{4.36} & 3.18 & 4.17 & 3.54 & 74.92
& 54.2 & 53.6 \\

Backbone-LLM-7B
& 31.41 / -- & 40.65 / --
& 78.67 / -- & 47.07 / -- & 42.18 / -- & --
& 79.33 & -- & --
& -- & -- & -- & -- & -- & --
& \textbf{66.5} & -- \\

\rowcolor{green!30} 
PRIME-Speech-9B
& 31.40 / \textbf{33.24} & \textbf{41.29} / \textbf{40.98}
& \textbf{79.00} / \textbf{74.42} & 46.98 / 44.54 & 42.04 / \textbf{40.18} & \textbf{3.33}
& \textbf{80.45} & \textbf{79.33} & \textbf{3.34}
& 68.9 & 4.23 & \textbf{3.37} & 4.36 & \textbf{4.29} & \textbf{78.76}
& 66.2 & \textbf{63.4} \\

\bottomrule
\end{tabular}}
\vspace{1mm}
\begin{minipage}{0.98\textwidth}
\footnotesize
$^\dagger$VocalNet is evaluated with the streaming mode from the official repository, not by first generating the complete text response and then synthesizing audio.
\end{minipage}
\end{table*}

%% file: sections/experiments.tex
\section{Results}
\label{sec:experiments}

\subsection{Baselines and Metrics}
\label{sec:exp_setup}
We compare PRIME-Speech with GLM-4-Voice~\cite{zeng2024glm}, Kimi-Audio~\cite{ding2025kimi}, Step-Audio-2-Mini~\cite{wu2025step}, VocalNet~\cite{wang2025vocalnet}, Qwen2.5-Omni~\cite{xu2025qwen2}, Qwen3-Omni~\cite{xu2025qwen3}, GPT-4o~\cite{gpt4o}, and Phi-4-MM-7B text backbone. Public systems use their recommended inference settings. Backbone-LLM is evaluated only in S2T mode because it has no speech-generation branch.

The evaluation covers speech translation on FLEURS~\cite{conneau2023fleurs} and CoVoST-2 X2EN~\cite{wang2021covost2}, spoken QA on UltraEval-Audio~\cite{shi2026ultraeval}, speech understanding on BigBench-Audio~\cite{srivastava2022beyond}, and conversational quality on VocalBench~\cite{liu2025vocalbench}. Since these benchmarks are mainly single-turn, we also construct a human-validated set of 28 conversations and 179 turns containing anaphora, ellipsis, and references to earlier responses.

We report S2T and S2S results separately. For S2S, Whisper Large-V3~\cite{radford2023robust} transcribes the generated speech before task scoring. Translation uses BLEU for S2T and ASR-BLEU for S2S; spoken QA, dialogue, and speech understanding use task accuracy. WER between the S2S transcript and the corresponding text response measures rendering consistency. VocalBench additionally reports UTMOS-based fluency as an automatic speech-naturalness metric. Unless otherwise stated, PRIME-Speech uses MTP horizon \(k{=}4\).

\input{tables/ablation}

\subsection{Main Results}
\label{sec:benchmarks_results}

Table~\ref{tab:speech_results_transposed} summarizes both S2T capability preservation and S2S performance. Across translation, spoken QA, multi-turn dialogue, and speech understanding, PRIME-Speech remains close to Backbone-LLM in S2T mode. Its S2S outputs retain strong task performance and show consistently low WER.

\subsubsection{Preserving the reasoning path}

The comparison with Backbone-LLM directly measures capability preservation. PRIME-Speech achieves nearly identical S2T scores on translation, spoken QA, and BigBench-Audio, showing that the added speech branch does not degrade the backbone's speech understanding or reasoning performance. PRIME-Speech also remains competitive with, and often outperforms, larger systems trained with substantially more data and computational resources. In particular, PRIME-Speech-9B surpasses Qwen3-Omni-30B on both S2S translation benchmarks and achieves much lower WER on spoken QA and multi-turn dialogue. It also obtains the best S2S scores on two of the three UltraEval-Audio tasks and on the multi-turn evaluation. These results show that strong speech-generation performance does not require sacrificing the capabilities of the original S2T model.

\subsubsection{From text correctness to spoken correctness}

The S2S results measure whether the spoken response preserves the content generated by the backbone. On translation and spoken QA, PRIME-Speech retains most of its S2T task performance after converting the response to speech. The multi-turn and BigBench-Audio results provide a stronger test because their responses depend on contextual understanding beyond transcript reconstruction. PRIME-Speech maintains small S2T--S2S gaps on both tasks, indicating that the audio branch preserves the backbone's decisions during speech generation. Task scores measure response correctness after transcription, while WER measures consistency between the spoken and text responses. Overall, PRIME-Speech combines competitive task performance with the lowest or among the lowest WER across all reported S2S tasks.


\subsection{Ablation Study}
\label{sec:ablation}
Table~\ref{tab:ablation_mtp} provides the controlled evidence behind the main results. It separates three effects that are often mixed in S2S systems: changing the reasoning backbone, improving the audio branch, and compressing codec-token generation.

\subsubsection{Why freeze the backbone}
The LoRA + Post LM and LoRA + ESI variants use the same data but update the backbone~\cite{hu2025slms2st, wu2025towards}. They remain viable on in-domain S2S metrics, but their lower BigBench-Audio S2T scores show the cost of adapting the reasoning model. This supports PRIME-Speech's decomposition: reasoning stays in the frozen backbone, while speech realization is learned in the post-decoder.

\subsubsection{What does MTP contribute}
MTP should be interpreted as an efficiency adapter, not as the source of semantic alignment. The comparison from the stage-1 model to the \(k{=}1\) model shows that the final audio curriculum already improves rendering quality even before reducing the codec rate. Increasing the horizon from \(k{=}1\) to \(k{=}4\) reduces the effective codec prediction rate from 25\,Hz to 6.25\,Hz while keeping S2S task scores broadly stable. The \(k{=}4\) setting preserves the main S2S behavior observed at smaller horizons, including strong UltraEval-Audio, multi-turn, and BigBench-Audio S2S performance. Thus MTP serves its intended role: compressing the codec-generation loop without changing the frozen reasoning path or causing significant task-quality loss.

\subsubsection{Multi-turn cache policy}
\input{tables/cache_reset}
Table~\ref{tab:cache_ablation} tests whether audio history should persist across assistant turns, with the key signal being the turn-wise trajectory rather than average WER. With text accumulation and turn-local audio reset, WER stays below 4\% in every turn-position bucket and S2S accuracy remains stable as dialogue length grows, indicating that the text-side cache preserves dialogue semantics while acoustic generation stays local to the current response.

Without audio reset, the first two turns appear acceptable, but stale audio states soon accumulate. From the third turn onward, WER jumps sharply and later exceeds 100\%, while S2S accuracy falls to zero in the \(\geq5\)-turn bucket. Because the text cache is accumulated in both rows, the degradation is isolated to reused audio-side state rather than lost dialogue memory. 
The ablation is therefore a causal diagnostic: audio KV and audio positions should reset at turn boundaries. 


\subsection{Efficiency Analysis}
\label{sec:efficiency}
\input{tables/efficiency}
Table~\ref{tab:mtp_efficiency} compares inference efficiency under the same hardware. VocalNet's shadow and dense talker gives high codec-generation throughput and low RTF, so it represents an efficiency-oriented point on the S2S design trade-off. However, Table~\ref{tab:speech_results_transposed} shows that it also has a larger S2T--S2S modality gap than PRIME-Speech. PRIME-Speech operates in a different regime: a 2B post-decoder is synchronized with hidden states from a frozen reasoning backbone, which costs more computation than a shallow talker at \(k{=}1\), but preserves the S2T pathway and yields stronger text--speech consistency.

Within this frozen-backbone regime, MTP provides a controllable efficiency knob. Increasing \(k\) from 1 to 4 lowers the effective codec rate from 25\,Hz to 6.25\,Hz, reduces TTFA from 1.07\,s to 0.39\,s, and reduces RTF from 1.088 to 0.296, while Table~\ref{tab:ablation_mtp} shows broadly stable S2S task quality. Thus our efficiency claim is not that PRIME-Speech dominates every lightweight talker in raw codec throughput, but that audio-side MTP substantially improves latency and RTF under the stronger constraint of preserving the frozen S2T backbone and reducing the S2T--S2S modality gap.


%% file: tables/ablation.tex
\begin{table*}[t]
\centering
\caption{Ablation on variants decoding pattern and multi-token prediction (MTP). ``Frame rate'' denotes the effective autoregressive decoding rate for codec-token generation (25\,Hz$/k$ for MTP=$k$). For UltraEval-Audio, values are reported as S2T / S2S.}
\label{tab:ablation_mtp}
\renewcommand{\arraystretch}{0.85}
\resizebox{\textwidth}{!}{
\begin{tabular}{l c cc cccc ccc cc}
\toprule
& & \multicolumn{2}{c}{FLEURS} & \multicolumn{4}{c}{UltraEval-Audio} & \multicolumn{3}{c}{Multi-turn} & \multicolumn{2}{c}{BigBench-Audio} \\
\cmidrule(lr){3-4}\cmidrule(lr){5-8}\cmidrule(lr){9-11}\cmidrule(lr){12-13}
\textbf{Variant} & \textbf{Frame rate} & \textbf{S2T / S2S} & \textbf{WER} &
\textbf{LLaMA-QA} & \textbf{TriviaQA} & \textbf{WebQ} & \textbf{WER} &
\textbf{S2T} & \textbf{S2S} & \textbf{WER} &
\textbf{S2T} & \textbf{S2S} \\
\midrule
LoRA + ESI~\cite{wu2025towards}         & 37.5Hz   & 29.37 / 31.11 & 2.07 & 70.00 / 66.33 & 45.81 / 43.76 & 40.51 / 39.18 & 3.25 & - & - & - & 53.75 & 53.25 \\
LoRA + Post LM~\cite{hu2025slms2st}     & 25Hz   & 29.59 / 30.96 & 2.62 & 70.33 / 67.00 & 45.32 / 42.79 & 40.41 / 38.25 & 5.13 & - & - & - & 52.96 & 52.36 
\\
\midrule
PRIME-Speech S1 Model         & 25Hz   & 31.39 / 33.57 & 1.51 & 79.00 / 73.33 & 46.98 / \textbf{45.71} & 42.04 / 39.43 & 6.12 & 80.45 & 79.21 & 3.67 & 66.30 & 59.10 \\
+ MTP=1          & 25Hz   & 31.39 / \textbf{33.58} & \textbf{1.45} & 79.00 / 72.33 & 46.98 / 45.03 & 42.04 / 40.07 & 5.66 & 80.45 & 79.33 & 3.61 & 66.30 & 63.86 \\
+ MTP=2          & 12.5Hz & 31.39 / 33.56 & 1.52 & 79.00 / \textbf{74.67} & 46.98 / 44.93 & 42.04 / \textbf{40.27} & \textbf{3.01} & 80.45 & \textbf{79.33} & \textbf{2.07} & 66.40 & \textbf{64.16} \\
+ MTP=4          & 6.25Hz & \textbf{31.40} / 33.24 & 2.19 & 79.00 / 74.42 & 46.98 / 44.54 & 42.04 / 40.18 & 3.33 & 80.45 & 79.33 & 3.34 & 66.20 & 63.38 \\
\bottomrule
\end{tabular}
}
\end{table*}

%% file: tables/cache_reset.tex
\begin{table}[t]
\centering
\caption{Multi-turn audio-cache ablation on the in-house set; each cell reports S2S accuracy / WER (\%) by turn-position bucket.}
\label{tab:cache_ablation}
\renewcommand{\arraystretch}{1.15}
\setlength{\tabcolsep}{4pt}
\resizebox{\columnwidth}{!}{
\begin{tabular}{lccccc}
\toprule
\multirow{3}{*}{\textbf{Cache policy}}
& \multicolumn{5}{c}{\textbf{Multi-turn S2S Acc.}$\uparrow$ / \textbf{WER}$\downarrow$ (\%)} \\
\cmidrule(lr){2-6}
& \textbf{1} & \textbf{2} & \textbf{3} & \textbf{4} & \textbf{$\geq$5} \\
\midrule
\rowcolor{green!20}
Text accum.\ + audio reset & 92.86 / 2.44 & 82.14 / 1.94 & 71.43 / 1.97 & 85.71 / 3.65 & 73.13 / 1.48 \\
\quad w/o audio reset & 92.86 / 2.77 & 78.57 / 5.62 & 39.29 / 65.57 & 10.71 / 129.63 & 0.00 / 143.27 \\
\bottomrule
\end{tabular}}
\end{table}

%% file: tables/efficiency.tex
\begin{table}[t]
\centering
\caption{Inference efficiency under different MTP horizons. Experiments are conducted on 1 NVIDIA H100 GPU.}
\label{tab:mtp_efficiency}
\renewcommand{\arraystretch}{0.85}
\setlength{\tabcolsep}{4pt}
\resizebox{\columnwidth}{!}{
\begin{tabular}{l c c c c c}
\toprule
\multirow{2}{*}{\textbf{System}} & \textbf{Frame rate} & \textbf{TTFT}$\downarrow$ & \textbf{TTFA}$\downarrow$ & \textbf{Throughput}$\uparrow$ & \multirow{2}{*}{\textbf{RTF}$\downarrow$} \\
 & (Hz) & (ms) & (s) & (tok/s) & \\
\midrule
Qwen2.5-Omni-7B & 50.0 & 58 & 1.01 & 45.75 & 1.093 \\
\midrule
VocalNet-8B ($k{=}1$) & 12.5 & 38 & 0.51 & 216.89 & 0.250 \\
\phantom{VocalNet-8B }($k{=}3$) & 6.25 & 38 & 0.40 & 220.16 & 0.243 \\
\phantom{VocalNet-8B }($k{=}5$) & 4.17 & 38 & 0.40 & \textbf225.93 & 0.243 \\
\midrule
PRIME-Speech ($k{=}1$) & 25.0 & 61 & 1.07 & 30.62 & 1.088 \\
\phantom{PRIME-Speech }($k{=}2$) & 12.5 & 60 & 0.63 & 62.17 & 0.548 \\
\phantom{PRIME-Speech }($k{=}4$) & 6.25 & 58 & \textbf{0.39} & \textbf{123.76} & \textbf{0.296} \\
\bottomrule
\end{tabular}}
\vspace{1mm}
\end{table}

%% file: sections/conclusion.tex
\section{Conclusion}
We presented PRIME-Speech, a framework for adding speech-to-speech capability to a strong S2T LLM without altering its reasoning backbone.
PRIME-Speech achieves strong results across speech-to-speech translation, spoken conversation, and speech understanding while maintaining a small S2T--S2S modality gap. Multi-token prediction further improves latency and real-time factor for high-rate codec tokens while preserving the overall S2S behavior of the frozen-backbone system.